\pdfoutput=1
\documentclass[11pt,twoside,a4paper,cmspaper,final,collab]{cms-tdr}
\def\svnVersion{cf4913d}\def\svnDate{2026/08/05}\def\cmsCernNoTag{CERN-EP-2026-047}\def\cmsCernDate{\today}\def\cmsMessage{Published in Physics Letters B as \href{http://dx.doi.org/10.1016/j.physletb.2026.140812}{\doi{10.1016/j.physletb.2026.140812}.}}
\begin{document}\cmsNoteHeader{B2G-22-001}

\newcommand{\TXbb}{\ensuremath{\mathrm{T}_\text{Xbb}}\xspace}
\newcommand{\TbqqPN}{\ensuremath{\mathrm{T}_{\Pb\Pq\Pq}^{\text{PNet}}}\xspace}
\newcommand{\TbqqDAK}{\ensuremath{\mathrm{T}_{\Pb\Pq\Pq}^{\text{DAK8}}}\xspace}
\newcommand{\pp}{\ensuremath{\Pp\Pp}\xspace}
\newcommand{\Tprime}{\HepParticle{T}{}{\prime}\xspace}
\newcommand{\bW}{\ensuremath{\PQb\PW}\xspace}
\newcommand{\tH}{\ensuremath{\PQt\PH}\xspace}
\newcommand{\tZ}{\ensuremath{\PQt\PZ}\xspace}
\newcommand{\tphi}{\ensuremath{\cPqt\PGf}\xspace}
\newcommand{\MTprime}{\ensuremath{m_{\Tprime}}\xspace}
\newcommand{\Mphi}{\ensuremath{m_{\PGf}}\xspace}
\newcommand{\mxmy}{\ensuremath{(\MTprime,\Mphi)}\xspace}
\newcommand{\MTprimeREC}{\ensuremath{m_{\Tprime}^{\text{rec}}}\xspace}
\newcommand{\MphiREC}{\ensuremath{m_{\PGf}^{\text{rec}}}\xspace}
\newcommand{\mxmyREC}{\ensuremath{(\MTprimeREC,\MphiREC)}\xspace}
\newcommand{\fb}{\ensuremath{\unit{fb}}\xspace}
\newcommand{\ttCR}{\ensuremath{\mathrm{CR}(\ttbar)}\xspace}
\newcommand{\qcdCR}{\ensuremath{\mathrm{CR}(\mathrm{QCD})}\xspace}
\newcommand{\SR}{\ensuremath{\mathrm{SR}}\xspace}
\newcommand{\ttCRF}{\ensuremath{\ttCR_{\mathrm{F}}}\xspace}
\newcommand{\qcdCRF}{\ensuremath{\qcdCR_{\mathrm{F}}}\xspace}
\newcommand{\ttCRP}{\ensuremath{\ttCR_{\mathrm{P}}}\xspace}
\newcommand{\qcdCRP}{\ensuremath{\qcdCR_{\mathrm{P}}}\xspace}
\newcommand{\SRF}{\ensuremath{\SR_{\mathrm{F}}}\xspace}
\newcommand{\SRP}{\ensuremath{\SR_{\mathrm{P}}}\xspace}
\newcommand{\particlenet}{\textsc{ParticleNet}\xspace}
\newcommand{\deepak}{\textsc{DeepAK8}\xspace}

\cmsNoteHeader{B2G-22-001}

\title{Search for a new heavy resonance decaying to a top quark and a neutral scalar boson in proton-proton collisions at \texorpdfstring{$\sqrt{s}=13\TeV$}{sqrt(s) = 13 TeV}}

\author*[cern]{The CMS Collaboration}

\date{\today}

\abstract{A first search at the LHC for a new heavy resonance decaying to a top quark and a neutral scalar boson \PGf in the fully hadronic final state is presented, where the \PGf boson is identified by its decay into a bottom quark-antiquark pair. The search is focused on final states in which the decay products of the highly Lorentz boosted top quark and \PGf boson are each reconstructed as a single, large-radius jet with distinct substructure. The analysis is performed using proton-proton collision data at a center-of-mass energy of 13\TeV, corresponding to an integrated luminosity of 138\fbinv, recorded with the CMS detector at the LHC in 2016--2018. The single production of a vector-like top quark, \Tprime, is used as a benchmark model for the signal process. The results of this search are combined with those of a previous CMS search in which semileptonic decays of the top quark were used. No significant excess of data is observed with respect to the background prediction. For the case where the neutral scalar is a standard model Higgs boson and the \Tprime quark width is 5\% of its mass, \Tprime quark masses between 0.85 and 1.3\TeV are excluded at 95\% confidence level and the most stringent limits to date are set for masses above 2\TeV. For other \PGf boson masses, upper limits as low as 0.1\unit{fb} are set on the product of the \Tprime quark production cross section and branching fraction for its decay to a top quark and a \PGf boson.}

\hypersetup{
    pdfauthor={CMS Collaboration},
    pdftitle={Search for a new heavy resonance decaying to a top quark and a scalar boson using proton-proton collisions at sqrt(s) = 13 TeV},
    pdfsubject={CMS},
    pdfkeywords={CMS, VLQ, boosted, scalar, jets, combination}
}

\maketitle

\section{Introduction}

The standard model (SM) of particle physics has been extensively tested and shown to be consistent with experimental results, yet it does not explain why the Higgs boson mass ($m_{\PH}$) lies near the electroweak (EW) scale, given the quadratic corrections arising from quantum loops. Naively, the Higgs boson mass would instead be expected to lie near the Planck scale, many orders of magnitude away from its measured value~\cite{Arkani-Hamed:1998jmv, Arkani-Hamed:2001nha, Agashe:2004rs, Panico:2015jxa}. Of the many beyond-the-SM (BSM) theories attempting to address this hierarchy problem, those of particular interest to the search reported in this Letter introduce new particles called vector-like quarks (VLQs), which are nonchiral, color-charged, spin-1/2 partners of the third-generation SM quarks~\cite{LHCsigVLQ}. 

The nonchiral nature of VLQs allows for Lagrangian mass terms that do not arise from Yukawa couplings to the Higgs field. Therefore, they are free from the stringent constraints on the existence of sequential fourth-generation quarks set by precision EW measurements at the Fermilab Tevatron and the CERN LHC~\cite{PhysRevD.86.013011}. At the LHC, VLQs can be produced singly or in pairs via the EW and strong interactions, respectively, with the former typically dominating for VLQ masses greater than 1\TeV. One such VLQ, the hypothetical top quark partner \Tprime, couples to the third-generation bottom (\PQb) and top (\PQt) quarks, and can decay to these quarks in association with a Higgs, \PW, or \PZ boson. In the case where the VLQ is a weak-isospin singlet, the branching fractions ($\mathcal{B}$) are 50, 25, and 25\% for the $\Tprime \to \bW, \, \Tprime \to \tH,$ and $\Tprime \to \tZ$ decays, respectively. For the doublet state, the branching fractions into \tH and \tZ are both 50\%~\cite{VLQhandbook}. Comprehensive reviews of searches for single and pair production of VLQs at $\sqrt{s}=13\TeV$ have been published by the ATLAS~\cite{ATLAS:2024fdw} and CMS~\cite{CMS:2024bni} Collaborations. These searches have targeted the direct decays of VLQs coupled predominantly to third-generation quarks and decaying through various bosonic channels. In particular, the ATLAS Collaboration has performed a combination of searches for singly produced vector-like top quark partners in the $\tH$ and $\tZ$ channels~\cite{ATLAS:2024xdc}.

However, the vector-like top quark partner need not couple solely to SM particles and, as a result, decays to new scalar or pseudoscalar particles become possible in nonminimal BSM scenarios~\cite{Fajfer_2013, Cacciapaglia:2019zmj, Benbrik:2019zdp, Aguilar-Saavedra:2019ghg, VLQtoScalar}. One such example is a two-Higgs-doublet model~\cite{Branco_2012} that is modified to include a VLQ~\cite{Benbrik:2019zdp}. The Higgs sector of such a model is extended to include an additional neutral scalar, neutral pseudoscalar, and two charged scalar bosons. The neutral scalar boson is predicted to decay to similar bosonic and fermionic final states as those of the SM Higgs boson, and the branching fraction of the VLQ decay to this scalar boson increases with the VLQ mass. 

Motivated by such models, this Letter presents a search for the EW production of a single top quark partner \Tprime in association with a bottom quark and decaying to a top quark and a neutral scalar boson \PGf, wherein the top quark decays hadronically and the \PGf boson is assumed to decay to a bottom quark-antiquark pair (\bbbar). To take advantage of the larger VLQ single production cross sections at high mass, we target \Tprime quark masses (\MTprime) ranging from 0.8 to 3\TeV and consider masses of the new scalar boson (\Mphi) ranging from 75 to 500\GeV. In this regime, the top quark and \PGf boson are each highly Lorentz boosted, such that their decay products merge, creating a dijet topology with large-radius jets in the final state.

Potential signal jet candidates are distinguished from background using neural network identification algorithms, referred to as ``taggers'', which discriminate the jets originating from top quark or Higgs boson decays from those produced via quantum chromodynamics (QCD), referred to henceforth as the QCD multijet background. Since the final state includes one jet from the hadronic decay of a top quark and one $\PGf \to \bbbar$ candidate jet, which could be a misidentified hadronic top quark decay, events involving top quark pair production (\ttbar) form the dominant background in this search. Although the selection on the tagging algorithm's outputs is efficient at removing QCD multijet events, these events nonetheless remain the next most dominant background, followed by minor contributions from \PW and \PZ boson production in association with jets, henceforth collectively termed {\PV}+jets.

We construct a binned likelihood model composed of QCD multijet, \ttbar, {\PV}+jets, and signal components in the plane spanned by the reconstructed mass of the scalar boson, \MphiREC, and the invariant mass of the system composed of the reconstructed scalar boson and top quark, \MTprimeREC. This analysis targets a localized excess of events in the two-dimensional (2D) plane of \mxmyREC that would correspond to the presence of a \Tprime quark resonance, decaying to a top quark and a \PGf boson. The QCD multijet background in the signal region (SR) of this analysis is modeled using the prediction from a signal-depleted control region (CR) in data defined by inverting the tagger selection on the \PGf candidate jet. All other background sources and the signal are modeled using Monte Carlo (MC) simulation. To avoid bias in the prediction of the QCD multijet background in the \SR, the data there are initially kept blind and the background estimation procedure is first validated using pseudo-data generated from the sum of the MC backgrounds and the QCD multijet background prediction obtained from multijet-enriched CR, as described in Section~\ref{sec:bkg_est}.

The results from a combination of the fully hadronic channel with the channel wherein the top quark decays semileptonically, described in detail in Ref.~\cite{semilep}, are presented. A charged-lepton veto in the fully hadronic channel ensures the orthogonality with the semileptonic channel by construction, and sources of systematic uncertainty that are common between the two channels are treated as correlated in the likelihood. Both analyses are performed using proton-proton (\pp) collision data collected with the CMS detector in 2016--2018 at $\sqrt{s}=13 \TeV$, corresponding to an integrated luminosity of 138\fbinv~\cite{CMS-LUM-17-003,CMS-PAS-LUM-20-001}. These are the first searches for the decay of a VLQ to a BSM particle.

This Letter is organized as follows. Section~\ref{sec:detector} provides an introduction to the CMS detector and event reconstruction. Section~\ref{sec:SigAndBkgs} describes the simulated signal and background samples used in this analysis. The event selection and object identification criteria are outlined in Section~\ref{sec:EventSelection}, followed by a description of the background estimation strategy in Section~\ref{sec:bkg_est}. Having thus introduced the search in the fully hadronic channel, a brief overview of the search performed in the semileptonic channel is provided in Section~\ref{sec:semileptonic}. Section~\ref{sec:SystUnc} summarizes the sources of systematic uncertainty in the analysis, as well as the treatment of the combination of the fully hadronic and semileptonic channels. Finally, the results are presented in Section~\ref{sec:Results} and this Letter is summarized in Section~\ref{sec:summary}. Tabulated results can be found in the HEPData record for this analysis~\cite{hepdata}.

\section{The CMS detector and event reconstruction}
\label{sec:detector}

The central feature of the CMS apparatus is a superconducting solenoid of 6\unit{m} internal diameter, providing a magnetic field of 3.8\unit{T}. Within the solenoid volume are a silicon pixel and strip tracker, a lead tungstate crystal electromagnetic calorimeter (ECAL), and a brass and scintillator hadron calorimeter, each composed of a barrel and two endcap sections. Forward calorimeters extend the pseudorapidity ($\eta$) coverage provided by the barrel and endcap detectors. Muons are measured in gas-ionization detectors embedded in the steel flux-return yoke outside the solenoid. A more detailed description of the CMS detector, together with a definition of the coordinate system used and the relevant kinematic variables, can be found in Ref.~\cite{CMS:2008xjf, CMS:2023gfb}.

Events of interest are initially selected using a two-tiered trigger system. The first level (L1), composed of custom hardware processors, uses information from the calorimeters and muon detectors to select events at a rate of around 100\unit{kHz} within a fixed latency of about 4\mus~\cite{CMS:2020cmk}. The second level, known as the high-level trigger, consists of a farm of processors running a version of the full event reconstruction software optimized for fast processing, and reduces the event rate to around 1\unit{kHz} before data storage~\cite{CMS:2016ngn, CMS:2024aqx}.

The particle-flow (PF) algorithm~\cite{CMS:2017yfk} aims to reconstruct and identify each individual particle in an event with an optimized combination of information from the various elements of the CMS detector. For each event, hadronic jets are clustered from these reconstructed particles (PF candidates) using the infrared- and collinear-safe anti-\kt algorithm~\cite{Cacciari:2008gp, Cacciari:2011ma}. Two different jet collections, referred to as AK4 and AK8 jets, are clustered using distance parameter values of 0.4 and 0.8, respectively. Jet momentum is determined as the vectorial sum of all particle momenta in the jet, and is found from simulation to be, on average, within 5--10\% of the true momentum over the whole transverse momentum (\pt) spectrum and detector acceptance. Additional \pp interactions within the same or nearby bunch crossings (pileup) can contribute additional tracks and calorimetric energy depositions to the jet momentum. The pileup per particle identification algorithm~\cite{Bertolini:2014bba, CMS:2020ebo} is used to mitigate the effect of pileup at the reconstructed particle level, making use of local shape information, event pileup properties, and tracking information. 

Jet energy corrections are derived from simulation to bring the measured response of jets to that of particle level jets on average. In situ measurements of the momentum balance in dijet, photon+jet, {\PZ}+jet, and multijet events are used to account for any residual differences in the jet energy scale between data and simulation~\cite{CMS:2016lmd}. The jet energy resolution amounts typically to 15--20\% at 30\GeV, 10\% at 100\GeV, and 5\% at 1\TeV~\cite{CMS:2016lmd}. Additional selection criteria are applied to each jet to remove jets potentially dominated by anomalous contributions from various subdetector components or reconstruction failures~\cite{CMS:2020ebo}. In the case of AK8 jets, masses are calculated after applying a modified mass-drop algorithm~\cite{Butterworth:2008iy, Dasgupta:2013ihk}, known as the ``soft-drop'' algorithm~\cite{Larkoski:2014wba}, to remove wide-angle soft and collinear radiation from the jets.

\section{Signal and background simulation}
\label{sec:SigAndBkgs}

The signal process $\pp \to \Tprime \to \PQt\PGf$ is generated at leading order for \Tprime quarks in a mass range of 0.8--3\TeV, in steps of 100\GeV, using the MC generator \MGvATNLO~\cite{Alwall:2014hca} (version 2.6.5). The \Tprime quark is treated as a weak-isospin singlet produced in association with a bottom quark, with a width of 1\% of its mass. Because the reconstructed mass of the \Tprime quark is insensitive to small variations in the generated width, the results can be compared with theoretical predictions for singlet \Tprime quarks with the decay widths of both 1\% and 5\% of their mass. We assume $\mathcal{B}(\PGf\to\bbbar)=100\%$ in the interpretation of the resulting cross sections in Section~\ref{sec:Results}, except for the special case where the scalar is the Higgs boson, $\Mphi=m_{\PH}=125\GeV$, in which case the results are scaled by the SM branching fraction for the decay $\PH\to\bbbar$. When interpreting the results in the context of the $\Tprime \to \PQt \PH$ decay, the cross sections are evaluated at next-to-leading order in the narrow-width approximation~\cite{Buchkremer:2013bha, Carvalho_2018, PhysRevD.101.115027} for singlet \Tprime VLQs with a fixed branching fraction of 25\% to \tH.

The \PGf boson is generated with masses ranging from 75 to 500\GeV in varying steps, with a minimum step size of 25\GeV for lower-mass signals, but increasing to 100\GeV at higher masses. In order to decrease these gaps in the signal MC sample coverage, we perform interpolation between generated signal mass points in the region where $\MTprime \gtrsim 1\TeV$ and $\Mphi < 350\GeV$ to produce signal shapes with 25\GeV spacing in \Mphi. To accomplish the interpolation, each generated signal mass point is first modeled by fitting a double-sided Crystal Ball function~\cite{CrystalBallRef, gaiser1982} to its reconstructed \MphiREC and \MTprimeREC spectra, which is then normalized to obtain a continuous probability distribution function representing the shape of the signal along each reconstructed mass axis. The shape of an interpolated signal of mass $M$, situated between two existing signals of masses $M_1$ and $M_2$, such that $M_1 < M < M_2$, is then calculated along both axes following the method described in Ref.~\cite{Read:1999kh}. The final shape of the interpolated signal is derived as the product of these two distributions, yielding a 2D distribution as a function of \mxmyREC. The normalization of the signal is obtained by linear interpolation between the expected yields of the existing signals of masses $M_1$ and $M_2$. The procedure is validated by running the algorithm on the generated signal mass points and comparing the resulting distributions to the original. For all these signal mass points, the differences are negligible and the algorithm successfully reconstructs the original signal shape. We use the same interpolation strategy to produce the shape and normalization of the systematic uncertainty templates for each interpolated signal.

The two main backgrounds in this search arise from \ttbar production in which the top quarks decay hadronically and from QCD multijet events. Other sources of background, such as {\PV}+jets production, are found to have minimal contributions. To simulate SM \ttbar production, the next-to-leading-order \POWHEG v2~\cite{Nason:2004rx, Frixione:2007vw, Powhegttbar, Alioli:2010xd, Re:2010bp} matrix element event generator is used, and for QCD multijet and {\PV}+jets simulation, \MGvATNLO~\cite{Alwall:2014hca} (versions 2.6.1 and 2.6.5, respectively) is used at leading order accuracy. The simulated QCD multijet samples are only used for checks of self-consistency of the background estimate and do not enter in the final likelihood model used to extract results.

Hadronization and parton showering are simulated using the \PYTHIA (version 8.240) software package~\cite{Sjostrand:2014zea}. The NNPDF3.1~\cite{NNPDF:2017mvq} parton distribution functions (PDFs) at next-to-next-to-leading order accuracy are used with the CP5~\cite{CP5} underlying event tune for all simulations. The CMS detector simulation is performed with \GEANTfour~\cite{GEANT4}.

To simulate the effect of pileup, additional simulated inelastic \pp collisions are superimposed on the hard scattering process using \PYTHIA. Simulated samples are reweighted to correct the pileup simulation by using the total inelastic cross section of 69.2~\unit{mb}~\cite{Aaboud:2016mmw,Sirunyan:2018nqx} to approximate the distribution of the number of pileup interactions in data, after which the simulated events are reconstructed in the same manner as the collision data.

\section{Event selection}
\label{sec:EventSelection}

Different sets of triggers based on jet properties are used for event selection across the three data-taking years. The first trigger criterion is based on the scalar \pt sum of all AK8 jets in the event (\HT), requiring an \HT of at least 900 (1050)\GeV for the 2016 (2017--2018) data set. An additional trigger used in the 2017 (2018) data set requires an event to possess $\HT>750$ (850)\GeV as well as an AK8 jet with a trimmed jet mass of at least 50\GeV. For these triggers, the jet mass is calculated by a trimming algorithm~\cite{Krohn2010-jn} similar to the soft-drop algorithm described in Section~\ref{sec:detector}. 

The second trigger criterion requires a single AK8 jet above a given \pt threshold in the event. In both 2017 and 2018 a trigger requiring a single AK8 jet with $\pt > 500\GeV$ was used. An additional trigger required an AK8 jet with $\pt > 400 \GeV$ together with a trimmed jet mass of at least 30\GeV. Such triggers with only a \pt criterion were not used in 2016. All these triggers are combined with a logical OR in order to maximize the trigger efficiency in each year. The trigger reaches full efficiency for a dijet invariant mass of approximately 1\TeV for the two leading AK8 jets, and simulated events with dijet invariant masses below this threshold are weighted according to trigger efficiency correction factors measured in an orthogonal data set.

The initial event selection focuses on exploiting the kinematic properties of the decays of very heavy resonances, which are produced nearly at rest, causing their decay products to be highly Lorentz boosted and reconstructed roughly opposite to one another in the transverse plane. Therefore, events are required to have at least two AK8 jets, each with $\pt > 350\GeV$, $\abs{\eta} < 2.4$, and a soft-drop jet mass of at least 50\GeV. The top quark and \PGf boson candidates are then chosen from the remaining jets as the two leading-\pt jets that have an azimuthal angular separation $\Delta\phi > \pi/2$. Additionally, all events must pass a set of standard data quality filters designed to remove events affected by noise in the various subdetectors~\cite{NoiseFilters}. Finally, all events containing reconstructed leptons satisfying loose identification criteria~\cite{CMS:2020uim,Sirunyan_2018} are vetoed in order to remain orthogonal to the selections imposed by the semileptonic channel~\cite{semilep}. 

After this initial selection, the two leading-\pt jets are then subject to identification as the $\PGf\to\bbbar$ (``\PGf jet'') and $\PQt\to \PQb\PW(\PQq\overline{\PQq}^\prime)$ (``\PQt jet'') candidates. Identification of the \PGf jet is performed using a discriminant formed by the output of the \particlenet~\cite{PhysRevD.101.056019} algorithm, referred to here as \TXbb. Identification of the \PQt jet is achieved using either the \particlenet discriminant, \TbqqPN, or the output of the \deepak~\cite{CMS:2020poo} algorithm, \TbqqDAK, depending on whether the event belongs to the \SR or the \ttbar-enriched \ttCR, respectively. A description of these algorithms and the selection criteria for the different analysis regions follows. 

Classification of the two jets as signal jet candidates is performed using the dynamic graph convolutional neural network \particlenet~\cite{PhysRevD.101.056019}. For every jet in an event, the \particlenet algorithm returns probability-like scores for the jet to have originated from the hadronic decay of some massive particle or from the hadronization of light quarks or gluons. In this search, the \particlenet ``top quark score'' is given by the algorithm's default output for discrimination between jets originating from the hadronic decay of a top quark and from the hadronization of lighter quarks or gluons, referred to here as \TbqqPN. The \particlenet ``\PGf score'' is given by a discriminant termed \TXbb, defined by $P(X_{\mathrm{MD}}\to\bbbar)/(P(X_{\mathrm{MD}}\to\bbbar) + P(\mathrm{QCD}))$, where $P(\mathrm{QCD})$ is the score corresponding to the QCD multijet background and $P(X_{\mathrm{MD}}\to\bbbar)$ is the probability-like score corresponding to a jet arising from the decay of some massive resonance to \bbbar. The subscript of the \PGf score indicates that it is mass decorrelated ($\mathrm{MD}$), meaning that the tagger response does not significantly change as a function of the jet mass, making this score criterion suitable to identify $\PGf\to\bbbar$ decays without biasing the background mass distributions. Mass decorrelation is achieved by enforcing a uniform distribution in the jet mass and \pt of the simulated signal and background samples used in the training of the network. Because the network is only trained on scalar resonances with masses ranging from 15 to 250\GeV, the selection efficiency for jets with \Mphi in this range is greater than those with masses above 250\GeV by a factor of four. 

Using these tagger scores, the dijet events are separated into one of three different analysis regions: a QCD multijet-enriched \qcdCR, a \ttbar-enriched \ttCR, and the \SR in which the search is performed. Because only two jets are considered in each event, selecting one jet as the \PQt jet candidate naturally selects the other as the \PGf jet candidate. To belong in the \SR, events must contain at least one jet passing the high-purity (HP) working point (WP) of the \particlenet top quark tagger corresponding to a 0.1\% mistagging rate, $\TbqqPN > \mathrm{HP~WP}$. If neither jet meets this criterion, the event is discarded. If only one jet passes the HP WP it is assigned as the \PQt jet candidate. If both jets pass, then the jet with the higher \TbqqPN score is assigned as the \PQt jet candidate. 

The \qcdCR is defined by inverting the tagging requirement on the \PQt jet, selecting only events in which the \PQt jet candidate has a score passing only a very loose tagger selection on \TbqqPN such that $0.2 < \TbqqPN < \mathrm{HP~WP}$. An additional \PGf tagging veto is placed on the \PQt jet candidate in this region, requiring that it has $\TXbb < 0.2$ to reduce possible signal contamination in the \qcdCR introduced by the \TbqqPN score inversion.

Following the identification of the \PQt jet candidate, additional \PGf tagging selection is imposed on the \PGf candidate jet in all events in both the \SR and \qcdCR, dividing both regions into ``Fail'' and ``Pass'' regions. These regions are defined identically in the \SR and \qcdCR, with the Fail region defined by \PGf candidate jets with $0.8 < \TXbb < 0.98$ and the Pass region by \PGf candidate jets with $\TXbb > 0.98$. 

A dedicated \ttCR is used to constrain the systematic uncertainties affecting the dominant \ttbar background. The \ttCR is defined similarly to the \SR at first, with one jet having $\TbqqPN > \mathrm{HP~WP}$. The second jet is then subject to a requirement of $\TXbb < 0.8$ to avoid overlap with events in the \SR. The Pass and Fail regions of the \ttCR are then populated by events where this second jet passes or fails the HP WP of the mass-decorrelated \deepak~\cite{CMS:2020poo} top quark tagger, \TbqqDAK, corresponding to a 0.1\% mistagging rate.

\section{Background estimation}
\label{sec:bkg_est}

A binned 2D likelihood model is constructed in the \mxmyREC plane from QCD multijet, \ttbar, {\PV}+jets, and signal components. The {\PV}+jets component appears mainly in the \qcdCR because of the inversion of the top quark tagging requirements. The \ttbar and {\PV}+jets backgrounds are estimated from MC simulation, for which the effects of systematic uncertainties are incorporated in the fit by altered distributions obtained from the simulated samples, corresponding to variations in a given uncertainty source. The QCD multijet component is instead estimated using a method in which the distribution of multijet events taken from data in the \TXbb Fail region is multiplied by a smooth, parametric transfer function (TF) to estimate the distribution in the \TXbb Pass region. Because the \TXbb tagger response is not correlated with the jet mass, the shape of the QCD multijet background in the Pass region can be obtained directly from the shape in the Fail region. The discussion in the following paragraph is intended to be generic for both the \SR and \ttCR, each of which has its own Pass and Fail tagging regions.

The TF is defined as the ratio of events passing to those failing the tagging requirement. For the \SR and \qcdCR this is the requirement on the second jet's \TXbb score, and for the \ttCR this is the requirement on the second jet's \TbqqDAK score. To estimate the QCD multijet background in the Pass tagging region, the QCD multijet event yield in every bin of the \mxmy plane of the Fail region ($n_{\mathrm{F}}^{\mathrm{QCD}}(i)$) is represented by an unconstrained parameter in the likelihood, whose initial value is given by the difference between the observed data ($n_{\mathrm{F}}^{\mathrm{Data}}(i)$) and the total background from MC simulation ($n_{\mathrm{F}}^{\text{MC bkgs}}(i)$) in that bin, prior to the fit. Thus, the number of QCD multijet events in a given bin $i$ of any Pass tagging region, $n_{\mathrm{P}}^{\mathrm{QCD}}(i)$, is given by
\begin{linenomath}
    \begin{equation}
        n_{\mathrm{P}}^{\mathrm{QCD}}(i) = n_{\mathrm{F}}^{\mathrm{QCD}}(i)\,\mathrm{TF}\mxmy, 
        \label{eq:QCDP_def}
    \end{equation}
\end{linenomath}
where 
\begin{linenomath}
    \begin{equation}
        n_{\mathrm{F}}^{\mathrm{QCD}}(i) = n_{\mathrm{F}}^{\text{Data}}(i) - n_{\mathrm{F}}^{\text{MC bkgs}}(i).
        \label{eq:QCDF_def}
    \end{equation}
\end{linenomath}
A simultaneous maximum likelihood fit of the Pass and Fail regions is performed to determine the optimal parameters for $n_{\mathrm{F}}^{\mathrm{QCD}}(i)$ and $\mathrm{TF}\mxmy$, which also ensures that the QCD multijet estimates in the Fail regions are not strictly equal to the difference between the data and simulated background yields.

The \qcdCR is used in the analysis only to validate the QCD multijet background estimation procedure described above, and to obtain an initial estimate for the TF in the \SR before the SR data are unblinded. The choice of parameterization for the TF was determined by Fisher tests~\cite{FisherFTest, Fisher} and found to be constant along both reconstructed mass axes in the \qcdCR. The postfit value of the constant TF, found to be $0.14 \pm 0.03$, was then used to generate pseudo-data in the \SR before unblinding to help further validate the QCD multijet estimation procedure. The pseudo-data generated by the application of the TF measured in the \qcdCR is found to be consistent with the expected QCD multijet background in the \TXbb Pass region of the \SR, as the fitted value of the TF in the \SR is $0.11\pm 0.02$. A separate TF was measured in the \ttCR whose best fit polynomial representation was also found to be constant along both reconstructed mass axes, with a fitted value of $0.010 \pm 0.001$. 

The total background in the \SR and \ttCR is then estimated by constructing a likelihood composed of the QCD multijet prediction and the \ttbar and {\PV}+jets contributions, which are represented by 2D histograms obtained from MC simulation and modified by the systematic uncertainties described in Section~\ref{sec:SystUnc}. The total background model, constructed as the sum of the individual background contributions using a Poisson model for each bin of the \mxmyREC distribution, is then used to test the signal hypothesis by comparing the number of events from the background-only and signal-plus-background models, where the various signal mass hypotheses enter the likelihood as 2D histograms obtained from MC simulation.

In order to constrain the systematic uncertainties affecting the dominant \ttbar backgrounds, the \SR and \ttCR are fit simultaneously. The QCD multijet background yields in both regions are estimated in parallel, but uncorrelated with one another, by using a separate TF for each region. The final TFs are determined to be constant factors for both the \SR and \ttCR.

\section{Summary of the semileptonic result used in the combination}
\label{sec:semileptonic}

A brief summary of the search strategy for the semileptonic top quark decay channel is presented in this Section, together with a description of the procedure for its combination with the fully hadronic channel. A complete description of the semileptonic analysis can be found in Ref.~\cite{semilep}.

The semileptonic analysis targets single production of a vector-like top quark partner decaying to a top quark and a new scalar boson where the top quark decays to a bottom quark, charged lepton ($\Pell$), and neutrino ($\nu$) via the $\PQt \to \PQb\PW \to \PQb\Pell\nu$ channel and the scalar boson decays to \bbbar. The analysis targets events with an AK4 jet and an AK8 jet containing the hadronization products of the \PQb quark and \PGf boson, one electron or muon, and the presence of missing transverse momentum, \ptvecmiss, defined as the negative vector \pt sum of all PF candidates in the event. 

Similarly to the hadronic channel, the semileptonic analysis uses the mass-decorrelated \particlenet \TXbb discriminant to identify boosted \PGf or Higgs boson candidates reconstructed as AK8 jets. Top quark candidates are reconstructed using the information from the charged-lepton candidate, one AK4 jet, and the \ptvecmiss, which is assumed to represent the neutrino \pt. A multiclass BDT algorithm is trained to distinguish genuine top quark candidates from the misidentified ones in QCD multijet events and all other types of background. The BDT output is used to define a top quark tagging discriminant used to distinguish the genuine from the misreconstructed top quark candidates.

Events are then categorized based on the loose and tight WP thresholds of the \TXbb and BDT top quark tagger scores. The loose and tight \TXbb tagger WP thresholds are identical to those used in the hadronic analysis and serve to separate events into the CR and SR, respectively, with the former used to check the closure of the background estimation procedure and to constrain the background uncertainties in the final maximum likelihood fit. Events are further categorized based on the lepton flavor and the presence or absence of at least one forward jet ($2.4 < \abs{\eta} < 4$), a proxy for the light quark produced in association with the \Tprime quark signal. A selection on \MphiREC of the \PGf boson candidate is made separately for each \PGf boson mass hypothesis, ensuring 95\% signal efficiency in each case. The signal strength is then extracted from a simultaneous maximum likelihood fit to the invariant mass spectrum of the top quark and \PGf boson candidates in all signal and control regions formed by this event categorization.

\section{Systematic uncertainties}
\label{sec:SystUnc}

The fully hadronic analysis takes into account several systematic uncertainties that can affect both the shape and normalization of the reconstructed mass distributions \mxmyREC in simulation, which are treated as nuisance parameters in the maximum likelihood fit used for signal extraction. The uncertainty in the integrated luminosity, which affects only the overall normalization of the simulated processes, is implemented as a nuisance parameter with a log-normal probability distribution. All other uncertainties also affect the shape of the mass distributions, and they are constructed such that the ${\pm}1$ standard deviation of each distribution is mapped to the ${\pm}1$ standard deviation of the corresponding unit Gaussian constraint. Unless mentioned otherwise, all sources of uncertainty described here are applied to both the histograms obtained from the generated MC signal samples, as well as the interpolated signal histograms obtained by the procedure described in Section~\ref{sec:SigAndBkgs}. 

The relative impacts of the different sources of systematic uncertainty on the best fit signal strength are reported here for a signal with $\MTprime=1100\GeV$ and $\Mphi=175\GeV$, corresponding to the mass hypothesis with the largest local significance of $1.4$ standard deviations. The dominant source of uncertainty in this search is the uncertainty in the TF used to derive the QCD multijet background estimate in the \SR, which represents a 12\% relative impact on the best fit signal strength. The next most dominant sources of uncertainty arise from the signal efficiency of the \TXbb and \TbqqPN taggers applied to the signal simulation, with relative impacts of 8\% each, as well as the misidentification efficiency of the \TXbb tagger applied to the \ttbar simulation in the \SR, with a relative impact of 10\%. 

We consider several sources of theoretical uncertainties related to the modeling of the simulated signal and background processes. The uncertainty in the PDFs used to generate the simulated samples is derived from the PDF variations provided by the {NNPDF3.1} set~\cite{NNPDF:2017mvq}. Uncertainties in the QCD renormalization and factorization scales are obtained by varying these scales independently by factors of 0.5 and 2 with respect to their default values, excluding the two extreme variations. 

Experimental uncertainties related to the detector conditions are also considered and applied to MC simulated samples for both signal and background processes. The uncertainty in the integrated luminosity of the data set, taken to be 1.6\%~\cite{CMS-LUM-17-003,CMS-PAS-LUM-20-001}, has a sub-percent impact on the results. The uncertainties in the jet energy scale and resolution are estimated as functions of \pt and $\eta$ of the jets~\cite{CMS:2016lmd} and correspond to a relative impact of 3--5\%. The uncertainty in the pileup description is evaluated by varying the total inelastic cross section by ${\pm}4.6\%$~\cite{Aaboud:2016mmw}.

During the 2016--2017 data-taking period, a gradual shift in the timing of the L1 trigger inputs from the ECAL in the $\abs{\eta} > 2.0$ region caused a trigger inefficiency, denoted as ``prefiring''~\cite{CMS:2020cmk}. The trigger missed between 10\% and 20\% of the events containing an electron (of $\pt \gtrsim 50\GeV$) or a jet (of $\pt \gtrsim 100\GeV$) in the $2.5 < \abs{\eta} < 3.0$ range, the exact loss depending on the \pt and $\eta$ of the trigger object. Correction factors were computed from data and applied to the acceptance evaluated by simulation. The uncertainties in these corrections are implemented as templates affecting the shape of the \pt and $\eta$ distributions for 2016 and 2017 simulated samples.

The trigger efficiency is measured in an orthogonal data set as a 2D function of \mxmyREC. A parametric function is fit to the 2D efficiency curve, which serves to model the efficiency turn-on while simultaneously reducing statistical fluctuations in the measurement. The MC simulation based distributions are then corrected by evaluating the fitted efficiency function at each event's reconstructed \mxmyREC invariant mass, and distributions corresponding to ${\pm}1$ standard deviations in the trigger efficiency are derived from the fit uncertainties.

A specific uncertainty in the \ttbar simulation arises from the top quark \pt reweighting procedure~\cite{Czakon:2017wor}, which is extrapolated to high \pt. This uncertainty accounts for the observed discrepancy between the \pt spectrum of \ttbar pairs in data and simulation, as measured by the ratio of the distributions of the top quark \pt in data and simulation as predicted by \POWHEG + \PYTHIA~\cite{CMS:2016oae, CMS:2018adi}. A data-to-simulation correction factor $w_t(\pt)$ is applied to simulated \ttbar events as a function of their \pt, which takes the form $w_t(\pt)=e^{c_1 0.0615 - c_2 \left(0.0005/\GeV\right)\pt}$, as obtained from Refs.~\cite{CMS:2016oae, CMS:2018adi}. The uncertainty is implemented as uncorrelated variations of ${\pm}50\%$ in the $c_1$ and $c_2$ parameters of the correction factor and has a 2--3\% relative impact on the best fit signal strength. 

Additionally, uncertainties in the jet tagging and mistagging efficiencies of the various taggers used in this analysis are treated as shape effects on the signal and dominant \ttbar background. The efficiency of the \TXbb tagger is calibrated in a data set enriched in jets arising from gluon splitting to \bbbar pairs, $\Pg\to\bbbar$, the characteristics of which have been sculpted to resemble those of a scalar using a boosted decision tree (BDT)~\cite{CMS:2025kje}, with \pt-dependent correction factors and their uncertainties applied to signal simulation as shape templates. To select a subset of the $\Pg\to\bbbar$ jets in a phase space that most closely resembles that of the signal, the BDT takes as input six variables related to the jet substructure, including information about secondary vertices (SVs) within the jet. An SV is a point in the detector where particle tracks converge that is displaced from the vertex corresponding to the hardest scattering in the event, called the primary vertex. Charged-particle tracks associated with an SV are used to calculate its kinematic properties, such as its invariant mass, \pt, and its transverse displacement relative to the primary vertex, $d_{xy}$. 

The six inputs to the BDT are the $N$-subjettiness ratio $\tau_{21}$~\cite{Thaler:2010tr}, the soft-drop masses of the two subjets, the \pt of the two SVs matched to each subjet, and the number of tracks associated with each SV. Proxy jets are then selected using nine thresholds of the BDT score that vary as functions of the \particlenet tagger discriminant score, such that the shape of the tagger discriminant is similar for both the proxy and signal jets. A fit to the mass of the SV with the maximum $d_{xy}$ significance, defined as $d_{xy}$ normalized by its uncertainty, is performed simultaneously for events with jets failing and passing the given \TXbb tagger WP. The tagging efficiency ratios between data and simulation obtained from the fit are then used to scale the MC sample yields to data. The calibration procedure yields data-to-simulation correction factors ranging from 0.9 to 1.3 for the \TXbb tagger efficiency, depending on the jet \pt and data-taking year. A comprehensive description of the calibration of the mass-decorrelated \particlenet algorithm is provided in Ref.~\cite{CMS:2025kje}.

The mistagging efficiency for the \TXbb tagger, as well as the tagging efficiencies for the \TbqqPN and \TbqqDAK discriminants, are calibrated in data using the tag-and-probe method~\cite{CMS:2011aa} with events enriched in \ttbar production. The \TXbb mistagging uncertainty is applied only to \ttbar MC samples, as top quark induced jets have a nonnegligible, \pt-dependent, probability of being misidentified by this tagger as a $\PGf\to\bbbar$ candidate. The data-to-simulation correction factors for the \TXbb mistagging efficiency range from 0.7 to 1.2, depending on the jet \pt and data-taking year. The uncertainty in the \TbqqDAK tagging efficiency is applied only to the \ttbar background in the \ttCR, which is the only analysis region where this tagger is employed.

Finally, the uncertainty in the QCD multijet background prediction in the \SR is derived during the simultaneous fit to data of the \SR and \ttCR. The uncertainty in the best fit TF parameter values enters the Hessian matrix used to calculate the total background uncertainty. The TF parameters are uncorrelated between the \SR and \ttCR and are treated as nuisance parameters with uniform probability distribution functions in the likelihood.

The fully hadronic and semileptonic searches entering the combination are statistically independent of one another because of a lepton veto applied in the fully hadronic channel, which rejects any event with isolated muons or electrons. For all common \MTprime and \Mphi hypotheses, a simultaneous maximum likelihood fit is performed to extract limits on the product of the cross section and branching fraction for the $\Tprime \to \PQt\PGf$ process, wherein systematic uncertainties that are common between the fully hadronic and semileptonic top quark decay channels are treated as fully correlated. These include uncertainties arising from reconstruction and calibration procedures such as the jet energy scale and resolution and the top quark \pt reweighting. Uncertainties in the \TXbb tagging and mistagging efficiency correction factors are correlated between the two analyses, since both use this tagger at the same WP for boosted \PGf boson identification. Theoretical uncertainties in both the renormalization and factorization scales as well as in the PDFs used in the generators are also taken to be correlated. Finally, uncertainties inherent to the accelerator and detector in both analyses are correlated, including those in the integrated luminosity and the L1 prefiring correction factors. Systematic uncertainties inherent to the QCD multijet background estimates are uncorrelated between the two channels, as these background estimates are performed in statistically independent regions. A detailed description of the uncertainty sources exclusive to the semileptonic channel, which include those arising from lepton identification and semileptonic top quark reconstruction, can be found in Ref.~\cite{semilep}.

\section{Results}
\label{sec:Results}

A 2D binned maximum likelihood fit to data of the \mxmyREC distributions is performed simultaneously for events in the \SR and \ttCR, using MC simulation to model \ttbar and {\PV}+jets, and control samples in data to predict the QCD multijet background. The inclusion of the \ttCR background model in the likelihood constrains the systematic uncertainties affecting \ttbar MC simulation, helping to model this dominant background in the \SR more accurately. The product of a data-driven estimate in the \TXbb Fail region and an unconstrained constant TF is used to predict the QCD multijet background yield in the \TXbb Pass region of the \SR, and the same estimation is performed in the \ttCR using a separate constant TF. 

\begin{figure*}[htbp!]
    \centering
    \includegraphics[width=0.49\textwidth]{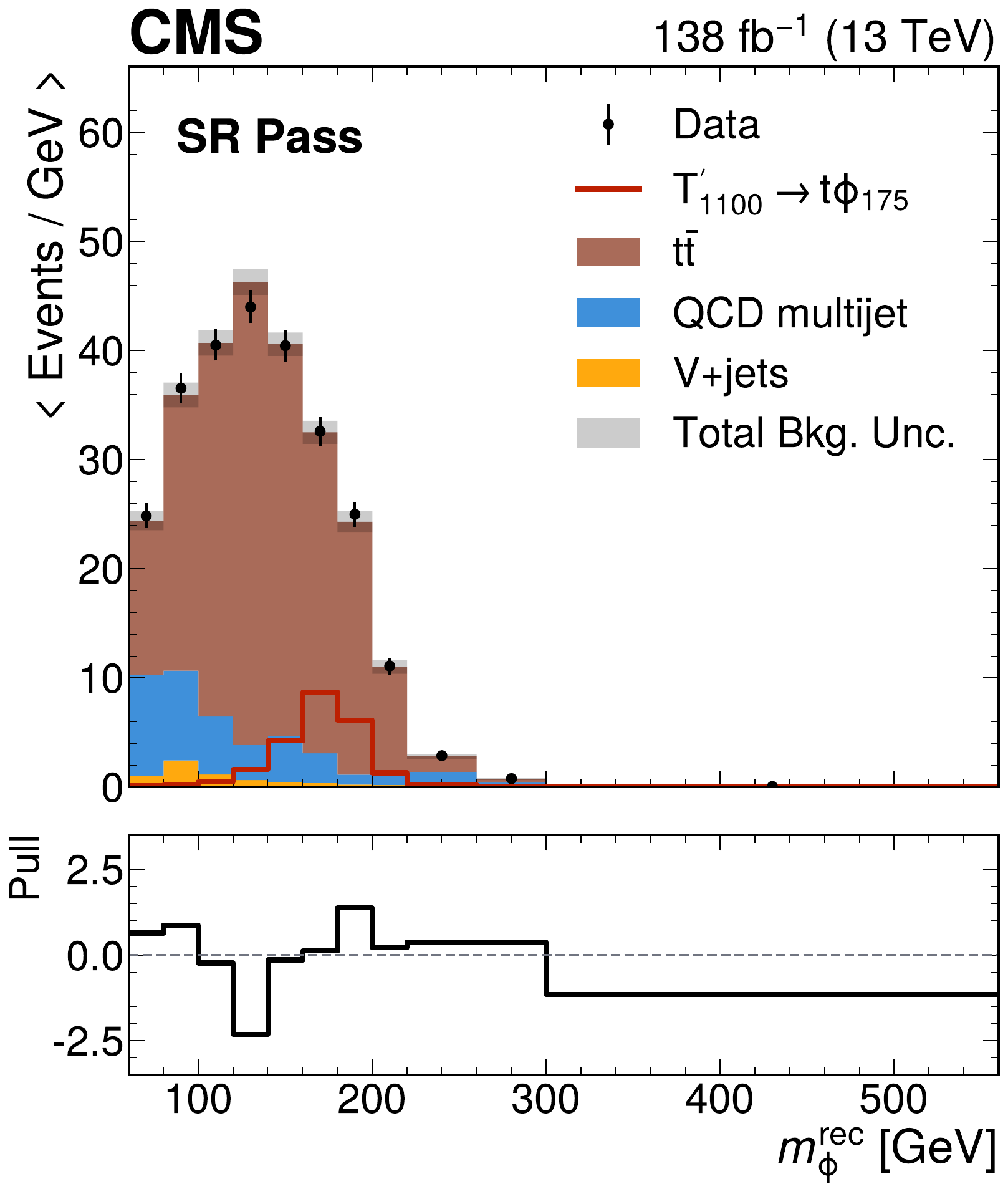}
    \includegraphics[width=0.49\textwidth]{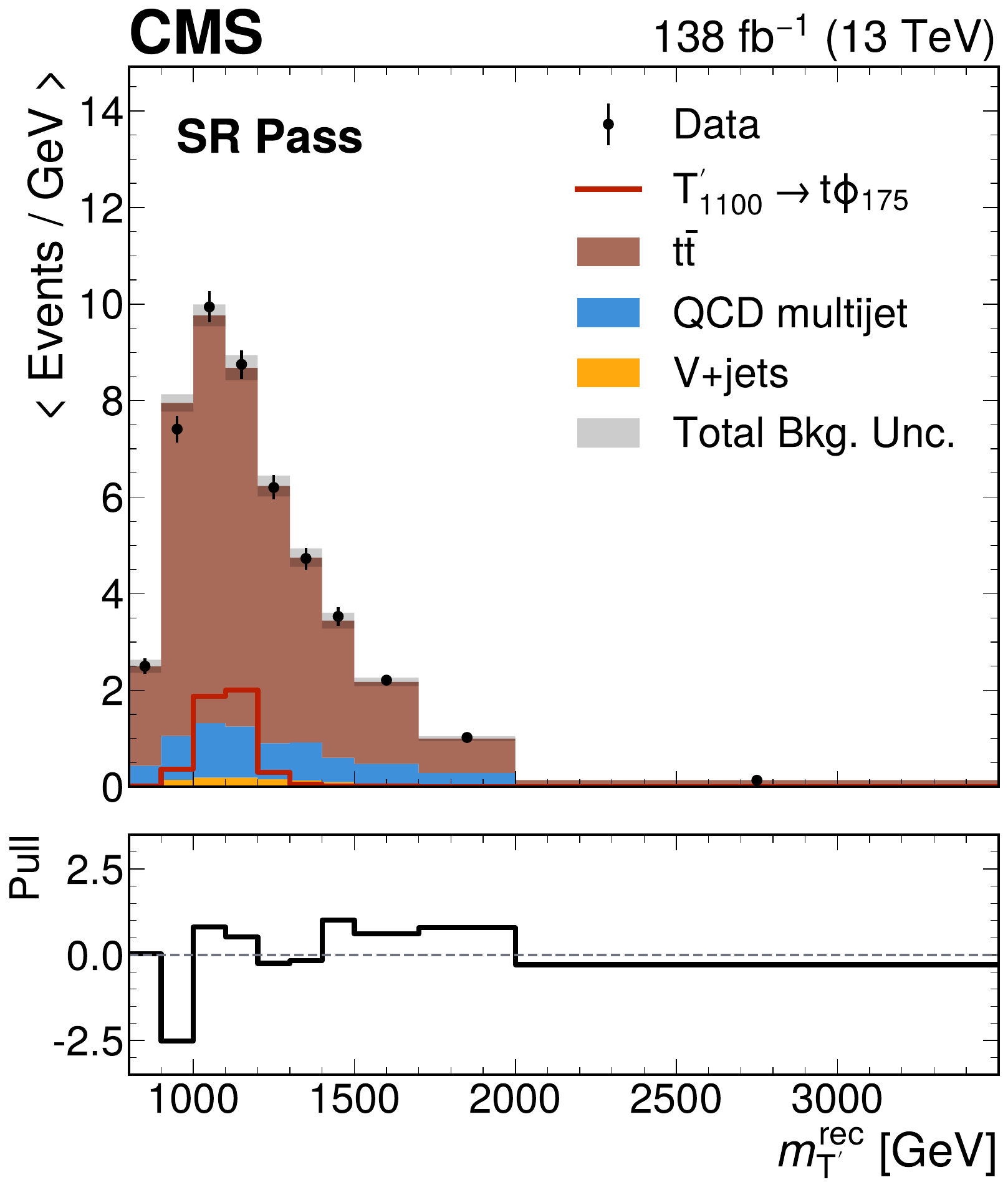}
    \caption{Postfit distributions of data and predicted background in the \SR under the background-only hypothesis, projected onto the \MphiREC (left) and \MTprimeREC (right) axes in the \TXbb Pass tagging region, \SRP. A $\Mphi=175\GeV,\, \MTprime=1100\GeV$ signal sample normalized to a cross section of $220\fb$ is overlaid for visualization. The $y$ axes of the top panels have been rescaled to display the number of events per bin, divided by the width of each bin.}
    \label{fig:postfit}
\end{figure*}

The results of the 2D fit to data in the hadronic channel are represented in Fig.~\ref{fig:postfit} as projections onto the reconstructed mass axes, and it is observed that the data are in agreement with the SM background prediction within uncertainties. A deficit in data relative to the predicted background is observed for the reconstructed \Tprime quark masses near 900\GeV and \PGf boson masses near 125\GeV. The deficit has a local (global) significance of 2.7 (1.5) standard deviations. The figure on the left shows the projection of the postfit distributions onto the \MphiREC axis and the one on the right the projection onto the \MTprimeREC axis. The lower panels show the pull distributions, defined for each bin as the difference between the observed data and the postfit background prediction divided by the square root of the difference between the squared Poisson uncertainty in the data and the squared postfit uncertainty in the total background estimate in that particular bin. 

The results from the fully hadronic channel are combined with those from the semileptonic channel in a global maximum likelihood fit using the CMS statistical analysis tool \textsc{Combine}~\cite{CMS:2024onh}, which is based on the \textsc{RooFit}~\cite{Verkerke:2003ir} and \textsc{RooStats}~\cite{Moneta:2010pm} frameworks. Shared sources of systematic uncertainty are correlated between the two channels in the fit and are listed in Section~\ref{sec:SystUnc}, while those exclusive to each channel are considered uncorrelated. The signal strength for each \MTprime hypothesis is extracted from a simultaneous fit of both channels' SRs, while utilizing their respective CRs to constrain the backgrounds. 

\begin{figure*}[htbp!]
    \centering
    \includegraphics[width=\textwidth]{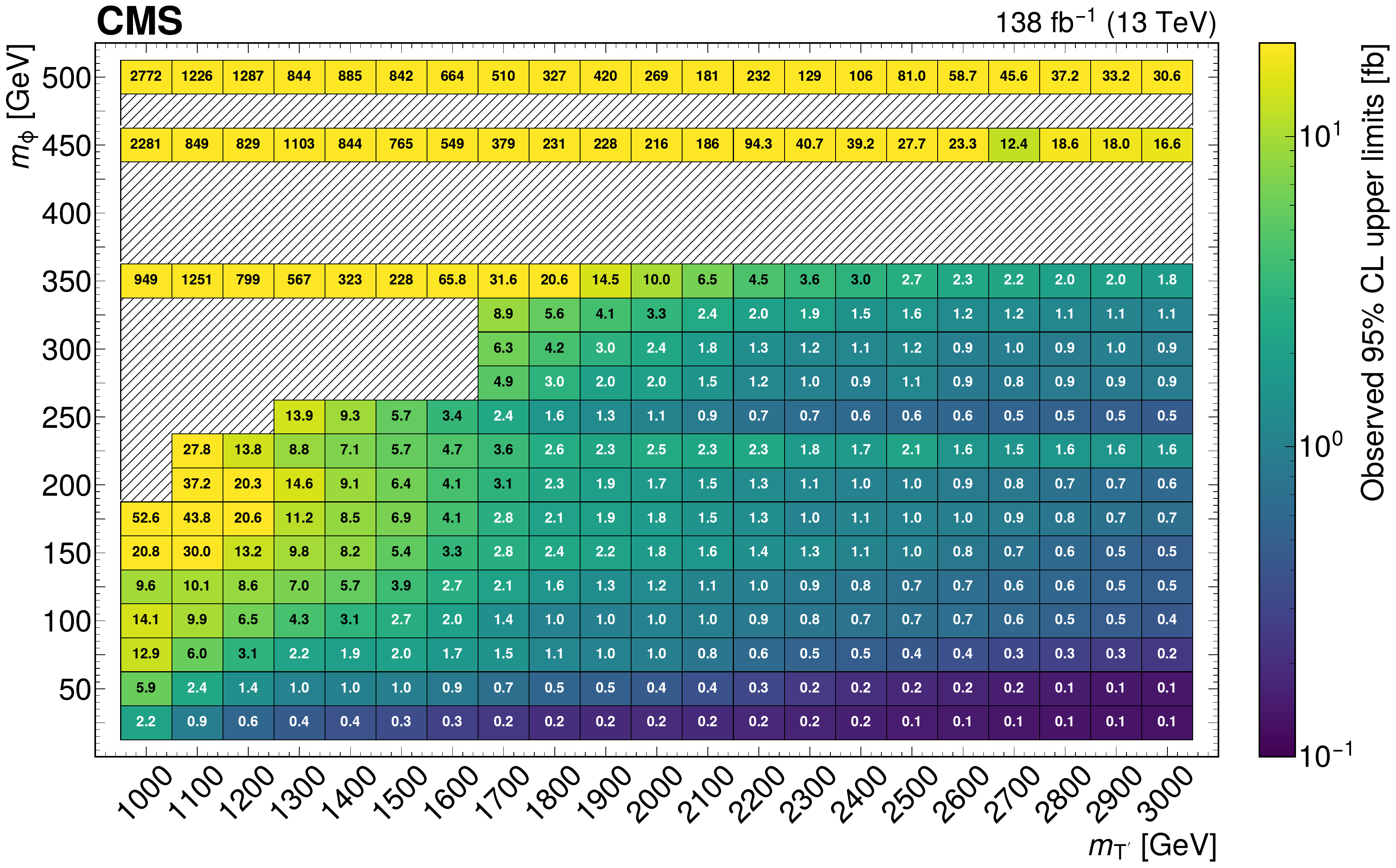}
    \caption{The upper limit at 95\% \CL on the product of cross section for the $\Tprime \to \PQt\PGf$ process and branching fraction $\mathcal{B}(\PQt\to \PQb\PQq\overline{\PQq}^\prime)$ as a function of $\mxmy$, assuming $\mathcal{B}(\PGf\to\bbbar)=100\%$. Masses of the $\PGf$ boson below 75\GeV are exclusive to the semileptonic channel, while $\Mphi>250\GeV$ are exclusive to the fully hadronic channel. Besides $\Mphi=225\GeV$, also exclusive to the fully hadronic channel, upper limits for $75 < \Mphi < 250\GeV$ are obtained from the combination of the two channels. Gaps in $\Mphi$ above 250\GeV are due to the limitations in the signal interpolation scheme in the hadronic channel, while gaps in $\Mphi < 250\GeV$ correspond to signals not considered by the semileptonic channel due to poor signal efficiency. All such gaps in coverage are represented by the hatched regions.}
    \label{fig:limits_2D}
\end{figure*}

The data are in agreement with the SM background prediction within uncertainties in both searches. The 95\% confidence level (\CL) expected and observed upper limits on the product of the cross section and branching fraction for \Tprime quark production are calculated using the \CLs criterion~\cite{CLS1, CLS2} with the profile likelihood ratio as the test statistic~\cite{CMS-NOTE-2011-005} using an asymptotic approximation~\cite{Cowan_2011}. The observed 95\% \CL upper limit is shown as a function of \mxmy in Fig.~\ref{fig:limits_2D}. No assumption is made for $\mathcal{B}(\Tprime\to\tphi)$, and $\mathcal{B}(\PGf\to\bbbar)$ is assumed to be 100\%. For $\MTprime < 1.5\TeV$, the combination achieves an improvement on the upper limits on the product of cross section and branching fraction for each channel by up to a factor of two for certain \Mphi. When interpreting the results in the case where the neutral scalar is the SM Higgs boson, the combination improves the mass exclusion limits by 100\GeV over the fully hadronic limits alone. For $\MTprime > 2\TeV$ the combination converges to the results obtained by the semileptonic channel, which is more sensitive primarily due to lower backgrounds. 

\begin{figure*}[htbp!]
    \centering
    \includegraphics[width=0.49\textwidth]{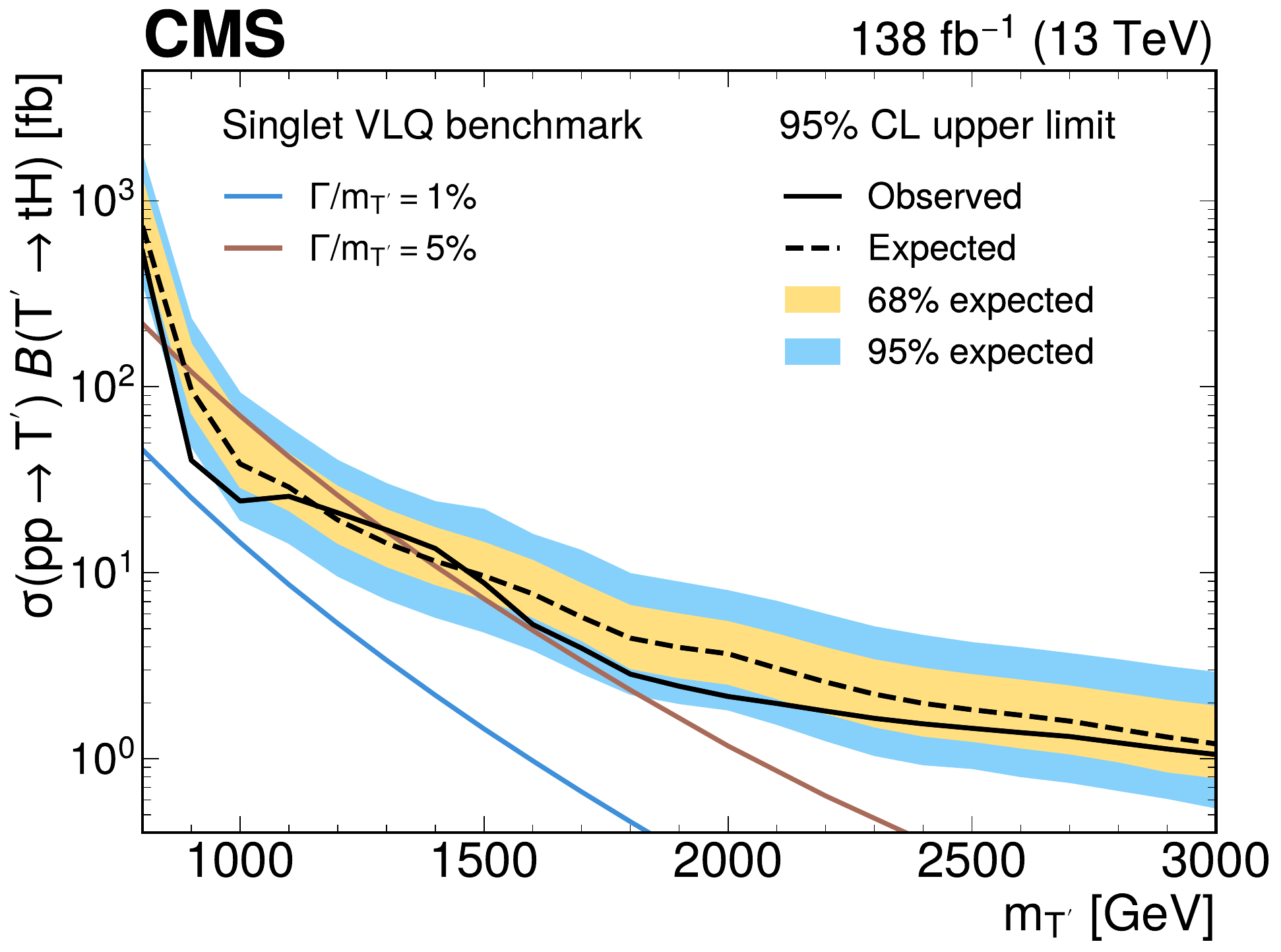}
    \includegraphics[width=0.49\textwidth]{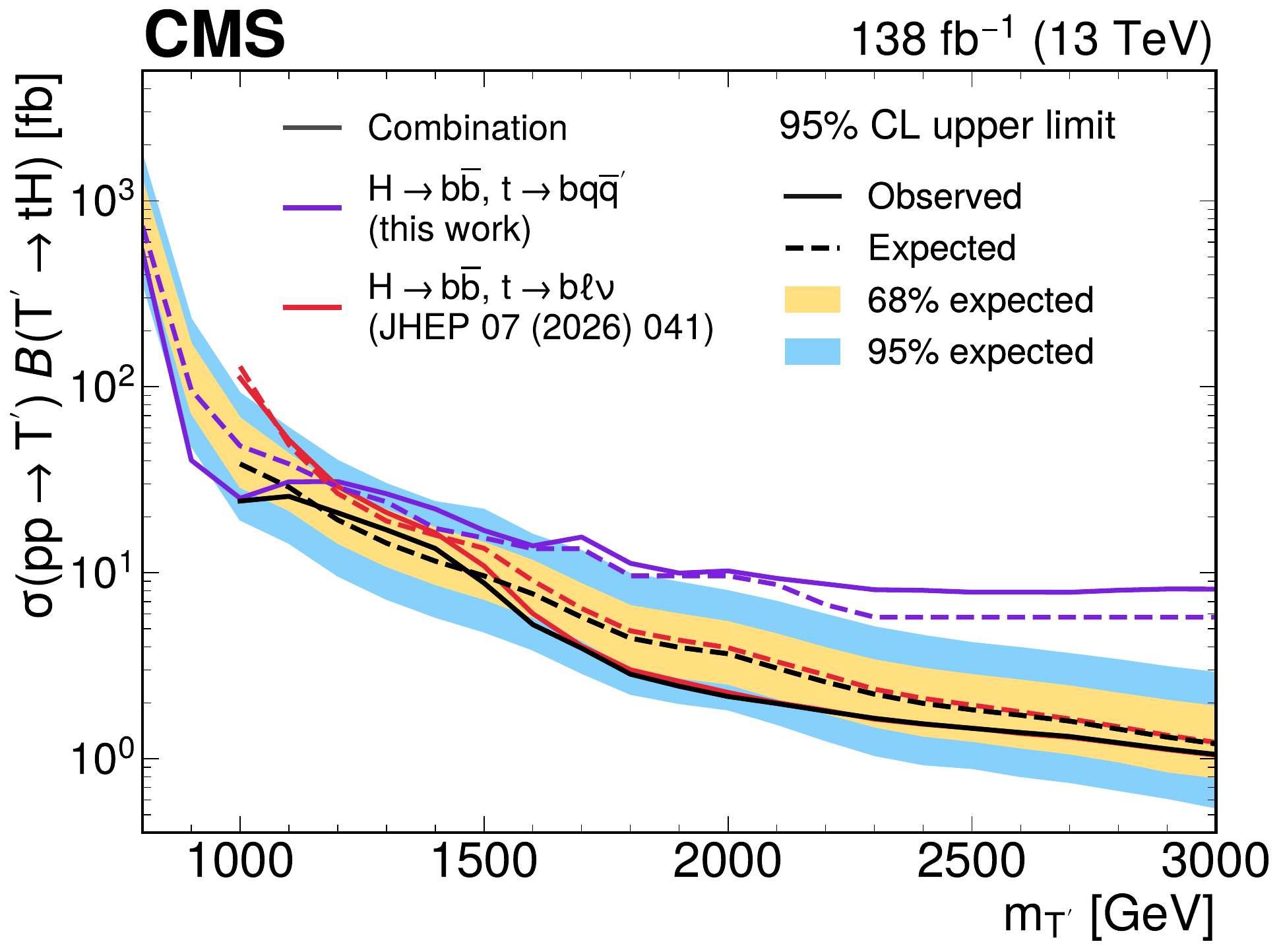}
    \caption{Upper limits at 95\% \CL on the product of the cross section and branching fraction for $\pp\to\Tprime\to\PQt\PH$ as functions of the \Tprime quark mass for fixed $\Mphi=m_{\PH}=125\GeV$. The limits differ from those in Fig.~\ref{fig:limits_2D} due to the assumption of the SM branching fraction for the $\PH\to\bbbar$ decay. In the left panel, the solid blue (brown) curves indicate the theoretical cross sections for the singlet \Tprime quark model assuming its width is 1\% (5\%) of its mass~\cite{Carvalho_2018, Buchkremer:2013bha}. The contributions to the combination from the semileptonic (red) and fully hadronic (purple) channels are detailed in the right panel, where it is shown that upper limits for \Tprime quark masses below 1\TeV are obtained from the fully hadronic channel only, while limits for all other mass points are obtained from the combination of the semileptonic and fully hadronic channels. The same holds for the shaded 68\% and 95\% expected upper limits in the right panel. In both panels, the median expected and the observed limits are depicted using dashed and solid lines, respectively.}
    \label{fig:limit_mH}
\end{figure*}

Limits at 95\% \CL on the product of the \Tprime quark production cross section and branching fraction are shown in Fig.~\ref{fig:limit_mH} as a function of \MTprime for $\Mphi=125\GeV$, corresponding to the SM Higgs boson. The limits for the two channels are shown alongside the results of the statistical combination. The results are interpreted here in the context of the singlet \Tprime quark, for resonances whose widths $\Gamma$ are 1 and 5\% of their mass. Masses of the \Tprime quark between 0.85--1.3\TeV are excluded at 95\% \CL for the scenario where the \Tprime quark width is 5\% of its mass, while no such exclusion can be made for the 1\% relative width scenario. Limits on the \Tprime quark coupling strength $\kappa$ as a function of the \Tprime quark mass can be found in Ref.~\cite{vlq-coupling-mass} for several different VLQ analyses in CMS alongside the results presented in this Letter. Here, $\kappa$ represents the strength of the coupling between the \Tprime quark, a standard model quark, and an EW gauge boson at the production vertex. 

\section{Summary}
\label{sec:summary}

A search for single production of a vector-like top quark partner \Tprime decaying to a top quark (\PQt) and a neutral scalar boson \PGf in the fully hadronic final state is presented, using proton-proton collision data recorded by the CMS experiment at $\sqrt{s}=13$ \TeV in 2016--2018 and corresponding to a total integrated luminosity of 138\fbinv. The hadronic decay products of the top quark and \PGf boson are expected to be highly Lorentz boosted from the decay of the massive \Tprime resonance, resulting in two large-radius jets in the final state. The results of this hadronic search are combined with those from a previous search in the semileptonic channels in a simultaneous maximum likelihood fit. No significant excess of data with respect to the background prediction is observed. Upper limits at 95\% confidence level (\CL) are set on the product of the production cross section and branching fraction for the decay $\Tprime\to \PQt\PGf$, representing the first results for the decay $\Tprime \to \PQt(\PQb\PQq\overline\PQq^\prime)\PGf(\PQb\overline\PQb)$ at the LHC. For the case where the neutral scalar is the standard model Higgs boson (\PH), upper limits are set on the product of the production cross section and the $\Tprime \to \PQt\PH$ branching fraction ranging from 23\fb for \Tprime quarks of mass $\MTprime=1\TeV$ to 1\fb at $\MTprime=3\TeV$. They exclude \Tprime quark masses of 0.85--1.3\TeV assuming the \Tprime quark is a weak-isospin singlet with a resonance width of 5\% of its mass, improving on all previous searches by the CMS Collaboration~\cite{CMS:2024bni} by up to a factor of three and representing the most stringent limits to date for masses above 2\TeV. For other \PGf boson masses, upper limits as low as 0.1\unit{fb} are set at 95\% \CL on the product of the \Tprime production cross section and branching fraction to $\PQt\PGf$.

\begin{acknowledgments}
We congratulate our colleagues in the CERN accelerator departments for the excellent performance of the LHC and thank the technical and administrative staffs at CERN and at other CMS institutes for their contributions to the success of the CMS effort. In addition, we gratefully acknowledge the computing centers and personnel of the Worldwide LHC Computing Grid and other centers for delivering so effectively the computing infrastructure essential to our analyses. Finally, we acknowledge the enduring support for the construction and operation of the LHC, the CMS detector, and the supporting computing infrastructure provided by the following funding agencies: SC (Armenia), BMBWF and FWF (Austria); FNRS and FWO (Belgium); CNPq, CAPES, FAPERJ, FAPERGS, and FAPESP (Brazil); MES and BNSF (Bulgaria); CERN; CAS, MoST, and NSFC (China); MINCIENCIAS (Colombia); MSES and CSF (Croatia); RIF (Cyprus); SENESCYT (Ecuador); ERC PRG and PSG, TARISTU24-TK10 and MoER TK202 (Estonia); Academy of Finland, MEC, and HIP (Finland); CEA and CNRS/IN2P3 (France); SRNSF (Georgia); BMFTR, DFG, and HGF (Germany); GSRI (Greece); MATE and NKFIH (Hungary); DAE and DST (India); IPM (Iran); SFI (Ireland); INFN (Italy); MSIT and NRF (Republic of Korea); MES (Latvia); LMTLT (Lithuania); MOE and UM (Malaysia); BUAP, CINVESTAV, CONACYT, LNS, SEP, and UASLP-FAI (Mexico); MOS (Montenegro); MBIE (New Zealand); PAEC (Pakistan); MSHE, NSC, and NAWA (Poland); FCT (Portugal); MESTD (Serbia); MICIU/AEI and PCTI (Spain); MOSTR (Sri Lanka); Swiss Funding Agencies (Switzerland); MST (Taipei); MHESI (Thailand); TUBITAK and TENMAK (T\"{u}rkiye); NASU (Ukraine); STFC (United Kingdom); DOE and NSF (USA).

\hyphenation{Rachada-pisek} Individuals have received support from the Marie-Curie program and the European Research Council and Horizon 2020 Grant, contract Nos.\ 675440, 724704, 752730, 758316, 765710, 824093, 101115353, 101002207, 101001205, and COST Action CA16108 (European Union); the Leventis Foundation; the Alfred P.\ Sloan Foundation; the Alexander von Humboldt Foundation; the Science Committee, project no. 22rl-037 (Armenia); the Fonds pour la Formation \`a la Recherche dans l'Industrie et dans l'Agriculture (FRIA) and Fonds voor Wetenschappelijk Onderzoek contract No. 1228724N (Belgium); the Beijing Municipal Science \& Technology Commission, No. Z191100007219010, the Fundamental Research Funds for the Central Universities, the Ministry of Science and Technology of China under Grant No. 2023YFA1605804, the Natural Science Foundation of China under Grant No. 12535004, and USTC Research Funds of the Double First-Class Initiative No.\ YD2030002017 (China); the Ministry of Education, Youth and Sports (MEYS) of the Czech Republic; the Shota Rustaveli National Science Foundation, grant FR-22-985 (Georgia); the Deutsche Forschungsgemeinschaft (DFG), among others, under Germany's Excellence Strategy -- EXC 2121 ``Quantum Universe" -- 390833306, and under project number 400140256 - GRK2497; the Hellenic Foundation for Research and Innovation (HFRI), Project Number 2288 (Greece); the Hungarian Academy of Sciences, the New National Excellence Program - \'UNKP, the NKFIH research grants K 131991, K 138136, K 143460, K 143477, K 147557, K 146913, K 146914, K 147048, TKP2021-NKTA-64, and 2025-1.1.5-NEMZ\_KI-2025-00004, and MATE KKP and KKPCs Research Excellence and Flagship Research Groups grants (Hungary); the Council of Science and Industrial Research, India; ICSC -- National Research Center for High Performance Computing, Big Data and Quantum Computing, FAIR -- Future Artificial Intelligence Research, and CUP I53D23001070006 (Mission 4 Component 1), funded by the NextGenerationEU program, the Italian Ministry of University and Research (MUR) under Bando PRIN 2022 -- CUP I53C24002390006, PRIN PRIMULA 2022RBYK7T (Italy); the Latvian Council of Science; the Ministry of Science and Higher Education, project no. 2022/WK/14, and the National Science Center, contracts Opus 2021/41/B/ST2/01369, 2021/43/B/ST2/01552, 2023/49/B/ST2/03273, and the NAWA contract BPN/PPO/2021/1/00011 (Poland); the Funda\c{c}\~ao para a Ci\^encia e a Tecnologia (Portugal); the National Priorities Research Program by Qatar National Research Fund;  MICIU/AEI/10.13039/501100011033, ERDF/EU, ``European Union NextGenerationEU/PRTR", projects PID2022-142604OB-C21, PID2022-139519OB-C21, PID2023-147706NB-I00, PID2023-148896NB-I00, PID2023-146983NB-I00, PID2023-147115NB-I00, PID2023-148418NB-C41, PID2023-148418NB-C42, PID2023-148418NB-C43, PID2023-148418NB-C44, PID2024-158190NB-C22, RYC2021-033305-I, RYC2024-048719-I, CNS2023-144781, CNS2024-154769 and Plan de Ciencia, Tecnolog{\'i}a e Innovaci{\'o}n de Asturias, Spain; the Chulalongkorn Academic into Its 2nd Century Project Advancement Project, the National Science, Research and Innovation Fund program IND\_FF\_68\_369\_2300\_097, and the Program Management Unit for Human Resources \& Institutional Development, Research and Innovation, grant B39G680009 (Thailand); the Eric \& Wendy Schmidt Fund for Strategic Innovation through the CERN Next Generation Triggers project under grant agreement number SIF-2023-004; the Kavli Foundation; the Nvidia Corporation; the SuperMicro Corporation; the Welch Foundation, contract C-1845; and the Weston Havens Foundation (USA).
\end{acknowledgments}

\bibliography{auto_generated}
\cleardoublepage \appendix\section{The CMS Collaboration \label{app:collab}}\begin{sloppypar}\hyphenpenalty=5000\widowpenalty=500\clubpenalty=5000\cmsinstitute{Yerevan Physics Institute, Yerevan, Armenia}
{\tolerance=6000
A.~Hayrapetyan, V.~Makarenko\cmsorcid{0000-0002-8406-8605}, A.~Tumasyan\cmsAuthorMark{1}\cmsorcid{0009-0000-0684-6742}
\par}
\cmsinstitute{Institut f\"{u}r Hochenergiephysik, Vienna, Austria}
{\tolerance=6000
W.~Adam\cmsorcid{0000-0001-9099-4341}, L.~Benato\cmsorcid{0000-0001-5135-7489}, T.~Bergauer\cmsorcid{0000-0002-5786-0293}, M.~Dragicevic\cmsorcid{0000-0003-1967-6783}, P.S.~Hussain\cmsorcid{0000-0002-4825-5278}, M.~Jeitler\cmsAuthorMark{2}\cmsorcid{0000-0002-5141-9560}, N.~Krammer\cmsorcid{0000-0002-0548-0985}, A.~Li\cmsorcid{0000-0002-4547-116X}, D.~Liko\cmsorcid{0000-0002-3380-473X}, M.~Matthewman, J.~Schieck\cmsAuthorMark{2}\cmsorcid{0000-0002-1058-8093}, R.~Sch\"{o}fbeck\cmsAuthorMark{2}\cmsorcid{0000-0002-2332-8784}, M.~Shooshtari\cmsorcid{0009-0004-8882-4887}, M.~Sonawane\cmsorcid{0000-0003-0510-7010}, N.~Van~Den~Bossche\cmsorcid{0000-0003-2973-4991}, W.~Waltenberger\cmsorcid{0000-0002-6215-7228}, C.-E.~Wulz\cmsAuthorMark{2}\cmsorcid{0000-0001-9226-5812}
% ...
\par}
\vskip\cmsinstskip
\dag:~Deceased\\
$^{1}$Also at Yerevan State University, Yerevan, Armenia\\
$^{2}$Also at TU Wien, Vienna, Austria\\
% ...
\end{sloppypar}
\end{document}